\documentclass[technote]{IEEEtran}
\ifCLASSINFOpdf
\else
\fi
\usepackage{graphicx} 
\usepackage{epsfig} 
\usepackage{amsmath} 
\usepackage{amssymb}  
\usepackage{psfrag}
\usepackage{color}
\usepackage{cite}

\graphicspath{{Figures/}}

\newcommand{\tR}{\mathcal{R}}
\newcommand{\tP}{\mathcal{P}}
\newcommand{\tQ}{\mathcal{Q}}
\newcommand{\tF}{\mathcal{F}}

\newtheorem{thm}{Theorem} 
 
\newtheorem{lem}{Lemma}
\newtheorem{coro}{Corollary}
\newtheorem{defi}{Definition}
\newtheorem{rem}{Remark}

\begin{document}
%
\title{Reduced-order Distributed Consensus Controller Design via Edge Dynamics}
%
%
%

\author{Dinh Hoa Nguyen, \IEEEmembership{Member, IEEE} 

\thanks{Dinh Hoa Nguyen is currently with Control System Laboratory, Department of Advanced Science and Technology, Toyota Technological Institute, 
        2-12-1 Hisakata, Tempaku-ku, Nagoya 468-8511, Japan. E-mail: hoadn.ac@gmail.com.} }

\newcounter{MYtempeqncnt}

\maketitle


\begin{abstract}
 
This paper proposes a novel approach to design fully distributed reduced-order consensus controllers for multi-agent systems (MASs) with identical general linear dynamics of agents. 
A new model namely edge dynamics representing the differences on connected agents' states is first presented. Then the distributed consensus controller design is shown to be equivalent to the synthesis of a distributed stabilizing controller for this edge dynamics. Consequently, based on LQR approach, the globally optimal and locally optimal distributed stabilizing controller designs are proposed, of which the locally optimal distributed stabilizing design for the edge dynamics results in a fully distributed consensus controller for the MAS with no conservative bound on the coupling strength. 
This approach is then further developed to obtain reduced-order distributed consensus controllers for linear MASs. 
Finally, a numerical example is introduced to demonstrate the theoretical results.

\end{abstract}



\section{Introduction}
%
%
%
%


 %

%
%

Multi-agent systems have gained much attention recently since there are a lot of practical applications, e.g.,     
power grids, wireless sensor networks, transportation networks, systems biology, etc, can be formulated, analyzed, and synthesized under the framework of MASs. 
A key feature in MASs is the achievement of a global objective by performing local measurement and control at each agent and simultaneously collaborating among agents using that local information.   

One of the most important and intensively investigated issues in MASs (and their applications) is the consensus problem due to its attraction in both theoretical and applied aspects 
\cite{Ren:2007,Olfati-Saber:2007}. The widely used consensus protocol \cite{Z.Li:2010} for identical general linear agents has the following form
\begin{equation}
	\label{consensus-law}
	u_{i}(t) = -\mu K\sum_{j\in\mathcal{N}_i}a_{ij}(x_i(t)-x_j(t)); i,j=1,\ldots,N,
\end{equation}
where $N$ is the number of agents, $u_{i}(t)$ is the control input vector of $i$th agent, $x_{i}(t)$ is the state vector, $\mu>0$ represents the coupling strength, $K$ is a coupling matrix, $\mathcal{N}_i$ is the neighboring set of $i$th agent, $a_{ij}>0$ are elements of the adjacency matrix (the definitions of graph-related terminologies will be given in Section \ref{pre}).

Consequently, a natural question arose for the consensus law (\ref{consensus-law}) is how to systematically design the coupling matrix $K$. 
One possibility is to employ the classical LQR optimal method of which $K$ is the solution of an algebraic Riccati equation, e.g. \cite{Cao:2010,Zhang:2011}. Nevertheless, \cite{Cao:2010} only considered agents with integrator dynamics while \cite{Zhang:2011} did not consider a global LQR performance index and hence the obtained distributed consensus controllers were locally optimal. The design of a globally optimal distributed consensus controller (\ref{consensus-law}) for general multi-input multi-output  (MIMO) linear MASs has recently  been presented in 
\cite{Movric:2014}. Unfortunately, all the obtained results hitherto for the consensus control law (\ref{consensus-law}) have conservative lower bounds for the coupling strength $\mu$. 

This paper firstly contributes a new approach to systematically design distributed consensus controllers for linear MASs consisting of MIMO agents whose communications are represented by undirected graphs. 
The key idea is to derive a transformed representation of MASs that we call {\it edge dynamics}. This can be considered as a generalized notion and development of the introduced 
``edge Laplacian" in \cite{Zelazo:2011} where only agents with single-integrator dynamics were treated. Based on the edge dynamics, we show the equivalence between the distributed consensus controller design for the initial MAS and the distributed stabilizing controller design for the edge dynamics. Subsequently, two different methods namely globally optimal and locally optimal designs are proposed to obtain such distributed stabilizing controller. 
The globally optimal stabilizing design leads to a lower bound condition for the coupling strength of the consensus law, which coincides with other studies in literature. 
However, {\it the interesting result for the locally optimal stabilizing synthesis reveals that the coupling strength can actually be arbitrarily small and positive such that the consensus is still achieved.} 
To the best of our knowledge, this result on the consensus design has not been achieved so far.

Next, the second contribution of this paper is on the design of reduced-order distributed consensus controllers. By further elaborating on the weighting matrices selection,  
the resulted LQR-based distributed consensus controller is shown to have a lower order than the previously obtained one. 
{\it Especially, if all agents have a single pole at the origin, we show that the MAS can achieve consensus with a first-order distributed controller. Furthermore, all eigenvalues of the closed loop MAS and the consensus speed can be explicitly analytically determined. }

The organization of next parts in the paper is as follows. 
Section~\ref{prob-form} introduces the edge dynamics and the design equivalence.  
Then systematic methods based on LQR approach to derive full-order and reduced-order distributed consensus controllers together with a numerical example are presented in Section~\ref{full-design} and Section~\ref{reduced-design}, respectively. 
Finally, some remarks are given in Section~\ref{concl}.

\section{Edge Dynamics and Design Equivalence}    
\label{prob-form}

\subsection{Preliminaries}
\label{pre}

Denote $\mathcal{G}$ the graph representing the information structure in a multi-agent system, 
where each node in $\mathcal{G}$ stands for an agent and each edge in $\mathcal{G}$ represents 
the interconnection between two agents. 
In this paper, we assume that $\mathcal{G}$ is undirected.  
Let us denote $E\in\mathbb{R}^{N\times M}$ the incidence matrix and $L$ the Laplacian matrix associated with $\mathcal{G}$, where $M$ is the number of edges in the graph. Subsequently, we have $L=EE^T$. 
Denote $L^{\dag}$ the generalized pseudoinverse of the graph Laplacian $L$ \cite{Gutman:2004}. 
The following lemma shows some useful properties of $E$ and $L^{\dag}$.
\begin{lem} \cite{Gutman:2004}
\label{pseudoinverse}
Suppose that $\mathcal{G}$ is connected then the following statements hold.
\begin{itemize}
	\item[(i)] $E^T\mathbf{1}_{N}=0$.
	\item[(ii)] $L^{\dag}L=LL^{\dag}=I_{N}-\frac{\mathbf{1}_{N}\mathbf{1}_{N}^T}{N}$.
\end{itemize} 
\end{lem}


Next, we introduce the notations used in this paper. The real, complex, complex with non-positive real part and complex with negative real part sets are denoted by $\mathbb{R}$, $\mathbb{C}$, $\mathbb{C}_{-}^{0}$, and $\mathbb{C}_{-}$, respectively. The superscript $^{*}$ denotes the complex conjugate transpose. 
Next, $\mathbf{1}_n$ represents the $n\times1$ vector with all elements equal to $1$, and $I_n$ denotes the $n\times n$ identity matrix. 
Moreover, $\succ$ and $\succeq$ denote the positive definite and positive semidefinite of matrices, respectively. 
On the other hand, $\mathrm{diag}\{\}$ represents diagonal or block-diagonal matrices, $\lambda_{\min}(A)$ denotes the eigenvalue with smallest, non-zero real part of $A$, 
and $\sigma(A)$ represents the eigenvalue set of $A$. 
Lastly, $\otimes$ stands for the Kronecker product.

\subsection{Edge Dynamics}

Consider an MAS consisting of $N$ identical agents whose model is described by
\begin{equation}
	\label{agent}
		\dot{x}_{i}(t) = Ax_{i}(t)+Bu_{i}(t), 
\end{equation}
where $A\in\mathbb{R}^{n\times n}$, $B\in\mathbb{R}^{n\times m}$; 
$x_{i}(t)\in\mathbb{R}^{n}$ and $u_{i}(t)\in\mathbb{R}^{m}$ are the state and input vectors of the $i$th agent, respectively.   
Initially, the agents are independent, i.e., the whole MAS can be represented by
\begin{equation}
	\label{mas}
		\dot{x}(t) = (I_{N}\otimes A)x(t)+(I_{N}\otimes B)u(t), 
\end{equation}
where 
$x(t)=[ x_1(t)^T, \ldots, x_N(t)^T]^T$, $u(t)=[ u_1(t)^T, \ldots, u_N(t)^T ]^T$. 
Then each agent collaborates with some of other agents to achieve a specific global objective namely consensus defined as follows.
\begin{defi} 
The agents (\ref{agent}) are said to reach a consensus if the following condition is satisfied,
\begin{equation}
	\label{consensus-defi}
	\lim_{t \rightarrow \infty}\|x_i(t)-x_j(t)\| = 0 ~\forall~i,j=1,\ldots,N.
\end{equation}
\end{defi} 
Hereafter, the time index $t$ is omitted in expressions of $x_{k},u_{k}$, and other variables 
just for the conciseness of mathematical representations. 

The following assumptions are employed in this paper. 
\begin{itemize}
	\item[{\bf A1:}] $\sigma(A) \in \mathbb{C}_{-}^{0}$, and at least one eigenvalue of $A$ is on the imaginary axis. 
	\item[{\bf A2:}] $(A,B)$ is controllable.
\end{itemize}
The purpose of assumption A1 is to avoid the consensus of agents to infinity or zero \cite{F.Xiao:2007} whereas assumption A2 is for the existence of a controller. 
Let us denote a new state vector and a new control input vector $$z\triangleq (E^T\otimes I_{n})x, ~w\triangleq (E^T\otimes I_{m})u,$$ then multiplying both sides of (\ref{mas}) with $E^T\otimes I_{n}$ gives us
\begin{align}
	\label{eq1}
	\dot{z} &= (E^T\otimes I_{n})(I_{N}\otimes A)x+(E^T\otimes I_{n})(I_{N}\otimes B)u \nonumber \\
	&= (E^T\otimes I_{n})[(L^{\dag}L+\frac{\mathbf{1}_{N}\mathbf{1}_{N}^T}{N})\otimes A]x+(E^T\otimes B)u \nonumber \\
	&= [(E^TL^{\dag}EE^T)\otimes A]x+(I_{M}\otimes B)w.
\end{align}
Consequently, Eq. (\ref{eq1}) can be further rewritten as follows,
\begin{equation}
	\label{edge-dyn}
	\dot{z} = [(E^TL^{\dag}E)\otimes A]z+(I_{M}\otimes B)w.
\end{equation}
We call the state-space representation (\ref{edge-dyn}) the {\it ``edge dynamics''} of the MAS (\ref{mas}). The state vector $z$ includes all the differences between the states of connected agents, i.e., it represents the MAS dynamics along the edges.  Therefore, the initial MAS is consensus if this edge dynamics is stabilized. 
If each agent is an integrator, i.e., $A=0,B=1$, (\ref{edge-dyn}) becomes
\begin{equation}
	\dot{z} = w ~\leftrightarrow~ \dot{x} = u, 
\end{equation}
which is much simpler than (\ref{edge-dyn}). 
The consensus protocol for this circumstance is $u=-\mu Lx$, which is equivalent to $w=-\mu L_{e}z$, where $L_{e}=E^TE$ is called the edge Laplacian \cite{Zelazo:2011}.  Hence, the closed-loop dynamics of the integrator MAS is
\begin{equation}
	\label{L-Le}
	\dot{z} = -\mu L_{e}z ~\leftrightarrow~ \dot{x} = -\mu Lx. 
\end{equation}
The equivalent transformation (\ref{L-Le}) shows us a direct connection between the original integrator MAS and its transformed dynamics along the graph edges as well as the connection between the Laplacian and the edge Laplacian of a graph. More details can be found in \cite{Zelazo:2011} but are omitted here since they are unused in our consensus control designs.   

Denote $\bar{L}=E^TL^{\dag}E$. The algebraic properties of $\bar{L}$ and $L_{e}$ are presented in the following lemma. 
\begin{lem}
\label{L-properties}
The following statements hold if $\mathcal{G}$ is connected.
\begin{itemize}
	\item[(i)] $L_{e}$ has exactly $N-1$ non-zero eigenvalues, which are equal to positive eigenvalues of $L$ while all other eigenvalues of $L_{e}$ if exist are $0$. 
	\item[(ii)] $\bar{L}$ has exactly $N-1$ non-zero eigenvalues, which are all equal to $1$, and other eigenvalues of $\bar{L}$ if exists are $0$.
\end{itemize}
\end{lem} 

\begin{IEEEproof}
(i): Let $(\lambda,v)$ be any eigen-pair of $L$ with $\lambda \neq 0$. 
Consequently, $\lambda E^Tv = E^TLv	= L_{e}E^Tv,$ 
hence $(\lambda,E^Tv)$ is also an eigen-pair of $L_{e}$. If $\mathcal{G}$ is connected then $L$ has a single zero eigenvalue while all other eigenvalues of $L$ are positive. 
Therefore, $L_{e}$ has $N-1$ non-zero eigenvalues, which are equal to non-zero eigenvalues of $L$. On the other hand, $\mathrm{trace}(L)=\mathrm{trace}(EE^T)=\mathrm{trace}(E^TE)=\mathrm{trace}(L_{e})$. Since $L_{e} \succeq 0$, all other eigenvalues of $L_{e}$ if exist must be $0$. 

(ii): Let $(\lambda,v)$ be any eigen-pair of $\bar{L}$. We have $\lambda Ev = E\bar{L}v = LL^{\dag}Ev = Ev.$ 
As a result, $\lambda=1$ is an eigenvalue of $\bar{L}$. Any other eigenvalue of $\bar{L}$ if exists must have the associated eigenvector $v$ satisfying $Ev=0$ and hence $\bar{L}v=0$ and $L_{e}v=0$, i.e., 
that eigenvalue is $0$. 
From the proof of part (i), the number of eigenvectors of $L_{e}$ associated with $0$ eigenvalues if exist is $M-N+1$. Thus, $\bar{L}$ has exactly $N-1$ eigenvalues equal to $1$ and all other eigenvalues if exist are $0$. 
\end{IEEEproof}

Let $U\in\mathbb{R}^{M\times M}$ be an orthogonal matrix that diagonalizes $\bar{L}$, and $\tilde{z} \triangleq (U^T\otimes I_{n})z$, $\tilde{w} \triangleq (U^T\otimes I_{m})w$. Subsequently, we obtain from (\ref{edge-dyn}) that
\begin{equation}
	\label{edge-dyn-1}
	\dot{\tilde{z}} = \left(\bar{\Gamma}\otimes A\right)\tilde{z}+(I_{M}\otimes B)\tilde{w},
\end{equation} 
where $\bar{\Gamma}=\mathrm{diag}\{0,I_{N-1}\}$ is a diagonal matrix including all eigenvalues of $\bar{L}$ in its diagonal. Next, let us partition $U$, the state and input vectors in (\ref{edge-dyn-1}) as follows,
\begin{equation}
	U = \begin{bmatrix} U_1 & U_2 \end{bmatrix},
	\tilde{z} = \begin{bmatrix} \tilde{z}_1 \\ \tilde{z}_2 \end{bmatrix},
	\tilde{w} = \begin{bmatrix} \tilde{w}_1 \\ \tilde{w}_2 \end{bmatrix},
\end{equation}
where $U_1\in\mathbb{R}^{M\times(M-N+1)}$, $U_2\in\mathbb{R}^{M\times(N-1)}$, $\tilde{z}_1\in\mathbb{R}^{n(M-N+1)}$, $\tilde{z}_2\in\mathbb{R}^{n(N-1)}$, $\tilde{w}_1\in\mathbb{R}^{n(M-N+1)}$, 
$\tilde{w}_2\in\mathbb{R}^{n(N-1)}$. Then (\ref{edge-dyn-1}) is equivalent to
\begin{equation}
	\label{edge-dyn-2}
	\begin{aligned}
		\dot{\tilde{z}}_1 &= (I_{M-N+1}\otimes B)\tilde{w}_1, \\
		\dot{\tilde{z}}_2 &= (I_{N-1}\otimes A)\tilde{z}_2+(I_{N-1}\otimes B)\tilde{w}_2, 
	\end{aligned}
\end{equation}
and $\tilde{z}_1=(U_1^T\otimes I_{n})z$, $\tilde{z}_2=(U_2^T\otimes I_{n})z$. 
Now, the initial MAS (\ref{mas}) is transformed along the edges to a system  composing of two independent subsystems. In the next section, we show that the consensus design for (\ref{mas}) is equivalent 
to the stabilizing design for the second subsystem in (\ref{edge-dyn-2}).

\subsection{Design equivalence}

In this paper, we assume that all states of agents are measurable hence a distributed state feedback consensus controller for the MAS (\ref{mas}) is designed. However, if not all states of agents can be measured then we can employ a decentralized Luenberger-type observer as in \cite{Nguyen:2015j} to obtain a distributed output feedback consensus controller. 
Due to space limitation, we do not duplicate the design of such decentralized Luenberger-type observer in the current paper.

Suppose that the stabilizing state feedback controller for the transformed edge dynamics (\ref{edge-dyn-2}) has the form $\tilde{w}=-\tF\tilde{z}$.  
Let $\Gamma \in \mathbb{R}^{(N-1)\times(N-1)}$ be the diagonal matrix including all non-zero eigenvalues of $L$ in its diagonal, and $V \in \mathbb{R}^{N\times N}$ is an orthogonal matrix such that 
\begin{equation}
	V^TLV = \begin{bmatrix} 0 & 0 \\ 0 & \Gamma \end{bmatrix}.
\end{equation}
Partitioning $V$ into $[V_1,V_2]$ where $V_1 \in \mathbb{R}^{N}$, $V_2 \in \mathbb{R}^{N\times(N-1)}$. Then 
\begin{equation}
	\label{V2-eq}
	LV_2=V_2\Gamma \Leftrightarrow V_2^TLV_2=\Gamma,
\end{equation}
since $V_2^TV_2=I_{N-1}$. 
The following theorem shows the equivalence between the distributed stabilizing controller for the transformed edge dynamics (\ref{edge-dyn-2}) and the distributed consensus controller for the initial MAS.
\begin{thm}
\label{ctlr-equi}
Suppose that $\mathcal{G}$ is connected. 
Let $U_2$ be chosen as $E^TV_2\Gamma^{-1/2}$, then the distributed stabilizing controller $\tilde{w}=-\tF\tilde{z}$ for the transformed edge dynamics with $\tF=F \otimes K$, 
$K \in \mathbb{R}^{m\times n}$ and 
\begin{equation}
	\label{F-form}
	F = \begin{bmatrix} 0 & 0 \\ 0 & \mu\Gamma \end{bmatrix},
\end{equation}
is equivalent to the consensus controller 
\begin{equation}
	\label{consensus-ctlr}
	u=-\mu(L \otimes K)x,
\end{equation}
for the initial MAS. Furthermore, $\tilde{z}_1(t)=0 ~\forall~ t\geq 0.$
\end{thm}

\begin{IEEEproof}
First, we show that the orthogonality of $U$ is satisfied with $U_2$ chosen to be $E^TV_2\Gamma^{-1/2}$. 
Indeed, $U_2^TU_1=\Gamma^{-1/2}V_2^TEU_1=0$ since $EU_1=0$ due to a fact that $E\bar{L}=E$. 
Furthermore, $U_2^TU_2=\Gamma^{-1/2}V_2^TEE^TV_2\Gamma^{-1/2}=\Gamma^{-1/2}\Gamma\Gamma^{-1/2}=I_{N-1}$. 
 
Next, multiplying to the left of (\ref{V2-eq}) with $E^T$ gives us
\begin{align}
	\label{eq:}
	L_{e}E^TV_2 &= E^TV_2\Gamma, \nonumber \\
	\Rightarrow L_{e}E^TV_2\Gamma^{-1/2} &= E^TV_2\Gamma^{1/2}, \nonumber \\
	\Rightarrow \Gamma^{-1/2}V_2^TEL_{e}E^TV_2\Gamma^{-1/2} &= \Gamma^{-1/2}V_2^TEE^TV_2\Gamma^{1/2}, \nonumber \\
	\Rightarrow U_2^TL_{e}U_2 &= \Gamma. 
\end{align}
On the other hand, $\bar{L}U_1=0$, which leads to $L_{e}U_1=0$ since $L_{e}\bar{L}=L_{e}$. Therefore, we obtain
\begin{equation}
	\label{F}
	F=\mu U^TL_{e}U.
\end{equation}
This is equivalent to
\begin{equation}
	\label{F-cond}
	F(U^TE^T) = \mu U^TL_{e}E^T = \mu U^TE^TL.
\end{equation}
Consequently, 
\begin{align}
	\label{equi}
	\tilde{w} &= -\tF\tilde{z}, \nonumber \\
	\Leftrightarrow  [(U^TE^T)\otimes I_{m}]u &= -(F \otimes K)[(U^TE^T)\otimes I_{n}]x, \nonumber \\
	\Leftrightarrow  [(U^TE^T)\otimes I_{m}]u &= -[(FU^TE^T)\otimes K]x, \nonumber \\
	\Leftrightarrow  [(U^TE^T)\otimes I_{m}]u &= -[\mu(U^TE^TL)\otimes K]x.
\end{align}
Since $U^TE^T = [EU_1, EU_2]^T = [0,V_2\Gamma^{1/2}]^T$, (\ref{equi})  occurs if and only if $[(\Gamma^{1/2}V_2^T)\otimes I_{m}]u = -[\mu(\Gamma^{1/2}V_2^TL)\otimes K]x$, which is equivalent to 
$[(V_{2}V_{2}^T)\otimes I_{m}]u = -[\mu(V_{2}V_2^TL)\otimes K]x$ by multiplying to the left with $(V_2\Gamma^{-1/2})\otimes I_{m}$.  
Note that $V_{1}=\frac{1}{\sqrt{N}}\mathbf{1}_{N}$, then $V_{2}V_{2}^T=I_{N}-V_{1}V_{1}^T=I_{N}-\frac{1}{N}\mathbf{1}_{N}\mathbf{1}_{N}^T$. Hence, we obtain 
$[(I_{N}-\frac{1}{N}\mathbf{1}_{N}\mathbf{1}_{N}^T)\otimes I_{m}]u = -[\mu([I_{N}-\frac{1}{N}\mathbf{1}_{N}\mathbf{1}_{N}^T]L)\otimes K]x=-\mu(L \otimes K)x$, since $\mathbf{1}_{N}^TL=0.$ 
This is equivalent to $u=-\mu(L \otimes K)x+(\mathbf{1}_{N}\otimes I_{m})u_{0}$ for any $u_{0} \in \mathbb{R}^{m}$. Since we are not interested in self-feedback inputs for agents, $u_{0}=0$ 
or equivalently $u=-\mu(L \otimes K)x$.  

On the other hand, we have $\tilde{z}_1=U_1^Tz=[(U_1^TE^T)\otimes I_n]x=0 $ since $EU_1=0$. 
\end{IEEEproof}

The interesting consequence of Theorem \ref{ctlr-equi} is that the distributed consensus design for (\ref{mas}) is equivalent to the synthesis of the distributed stabilizing controller 
$\tilde{w}_2=-(\mu\Gamma \otimes K)\tilde{z}_2$ for the transformed edge dynamics, which even has lower dimension.

\begin{rem}
If $\mathcal{G}$ is a spanning tree then $M=N-1$ and hence $\bar{L}=I_{N-1}$. Then we do not need the additional transformation (\ref{edge-dyn-1}).  
Therefore, all results here and in subsequent sections are derived with $\tilde{w}_{2}$ and $\tilde{z}_{2}$ replaced by $w$ and $z$, respectively. 
\end{rem}

\section{Full-Order Distributed Consensus Design}  
\label{full-design}

\subsection{Globally Optimal Stabilizing Design for Edge Dynamics}  
\label{stab-design}


Consider a performance index 
\begin{equation}
	\label{p-index}
	{\cal J} = \int_{0}^{\infty}{(\tilde{z}_2^T\tQ \tilde{z}_2+\tilde{w}_2^T\tR \tilde{w}_2)dt},
\end{equation}
where $\tQ\in\mathbb{R}^{(N-1)n \times (N-1)n}$, $\tQ\succeq 0$, $\tR\in\mathbb{R}^{(N-1)m \times (N-1)m}$, $\tR\succ 0$ are the weighting matrices. 
Subsequently, the stabilizing state feedback controller is $\tilde{w}_2=-F_2\tilde{z}_2$ with 
$F_2=\tR^{-1}(I_{N-1}\otimes B)^T\tP$ where $\tP$ is the positive definite solution of the following global Riccati equation
\begin{align}
	\label{Req}
	0 =& \, \tP[I_{N-1}\otimes A]+[I_{N-1}\otimes A]^T\tP+\tQ  \nonumber \\
	  & -\tP(I_{N-1}\otimes B)\tR^{-1}(I_{N-1}\otimes B)^T\tP.
\end{align}
It can be seen that (\ref{Req}) is a distributed structured Riccati equation and its solution $\tP$ should also have a distributed structure such that the stabilizing feedback gain $F_2$ has a specific form shown in Theorem \ref{ctlr-equi}. 
In the literature, there have been some studies on structured LQR designs for MASs, e.g. \cite{Motee:2008,Tsubakino:2013,Smith:2005}, which assumed that the system matrices and the weighting matrices belong to some special classes, e.g. operator algebras or semigroups so that the solution of Riccati equation and the feedback gain also have that structure. Nevertheless, we do not need those assumptions in our work. 
Therefore, our design is applicable for a broader class of systems. 

The following theorem reveals our design for distributed globally optimal stabilizing LQR controllers.
\begin{thm}
\label{stabilizing-ctlr}
Let the weighting matrices be selected as follows,
\begin{equation}
	\label{w-matrices}
	\begin{aligned}
		\tQ &= I_{N-1}\otimes Q_1+(\mu\Gamma-I_{N-1})\otimes Q_2, \\
		\tR^{-1} &= \mu\Gamma \otimes R_1, 
	\end{aligned}
\end{equation}
where $Q_1\in\mathbb{R}^{n\times n},Q_2\in\mathbb{R}^{n\times n}$, $Q_1\succeq 0$, $R_1\in\mathbb{R}^{m\times m}$, $R_1\succ 0$, $Q_2=P_1BR_1B^TP_1$, such that $(\tQ^{1/2},I_{N-1}\otimes A)$ is detectable, 
where $P_1$ is the solution of the following local Riccati equation
\begin{equation}
	\label{Req-local}
	P_1A+A^TP_1+Q_1-P_1BR_1B^TP_1=0.
\end{equation}
Moreover,  
\begin{equation}
	\label{mu}
	\mu \geq \frac{1}{\lambda_{\min}(L)}.
\end{equation}
Then the unique solution of the Riccati equation (\ref{Req}) is $\tP=I_{N-1}\otimes P_1$. 
Furthermore, the optimal stabilizing feedback controller gain for the transformed edge dynamics is
\begin{equation}
	\label{transformed-ctlr}
	F_{2}=\mu\Gamma\otimes(R_1B^TP_1).
\end{equation}
\end{thm}

\begin{IEEEproof}
With the weighting matrices selected in (\ref{w-matrices}) and $\mu$ chosen in (\ref{mu}), $\tQ \succeq 0$ and $\tR \succ 0$. 
Note that $(\mathcal{Q}^{1/2}, I_{N-1}\otimes A)$ is detectable if $(Q_{1}^{1/2},A)$ is detectable since it means $(I_{N-1}\otimes Q_{1}^{1/2}, I_{N-1}\otimes A)$ is also detectable and adding the term 
$(\mu\Gamma-I_{N-1})\otimes Q_{2} \succeq 0$  does not break its detectability.
Then the global Riccati equation (\ref{Req}) can be rewritten as follows,
\begin{align}
	\label{eq2}
	0 =& \tP(I_{N-1}\otimes A)+(I_{N-1}\otimes A)^T\tP+I_{N-1}\otimes Q_1 \nonumber \\
	&  +(\mu\Gamma-I_{N-1})\otimes Q_2-\tP[\mu\Gamma \otimes(BR_1B^T)]\tP. 
\end{align}
Consequently, we can observe that (\ref{eq2}) is satisfied if $\tP=I_{N}\otimes P_1$,   
$Q_2=P_1BR_1B^TP_1$, and $P_1$ is the solution of the local Riccati equation (\ref{Req-local}). 
Thus, $\tP=I_{N}\otimes P_1$ is a solution of (\ref{eq2}). 
On the other hand, the global Riccati equation (\ref{Req}) has a unique solution, therefore $\tP=I_{N}\otimes P_1$ is indeed that unique solution.
As a result, the optimal stabilizing feedback controller gain for the edge dynamics can be calculated as follows,
\begin{equation*}
	F_{2} = \tR^{-1}(I_{N-1}\otimes B)^T(I_{N-1}\otimes P_1)= \mu\Gamma \otimes(R_1B^TP_1). 
\end{equation*}
\end{IEEEproof}

\begin{rem}
The optimal performance index in this case is
\begin{equation}
	\label{J-global}
	{\cal J}_{\min} = \tilde{z}_{2}(0)^T\tP \tilde{z}_{2}(0) = \sum_{i=1}^{N-1}\tilde{z}_{2,i}(0)^TP_1\tilde{z}_{2,i}(0),
\end{equation} 
which is 
independent of the information graph, where $\tilde{z}_{2}(0)=[\tilde{z}_{2,1}(0)^T,\ldots,\tilde{z}_{2,N-1}(0)^T]^T$ is the vector of initial conditions. Therefore, the cost for stabilization is not changed even if the communication structure in the MAS is varied. This is more advantageous than other studies, e.g. \cite{Movric:2014}, where the minimum performance index depends on the communication graph.   
\end{rem}

\begin{rem}
\label{coupling-bound}
The condition for the coupling strength $\mu$ in Theorem \ref{stabilizing-ctlr} is similar to other existing results \cite{Z.Li:2010,Ma:2010,F.Xiao:2007}. However, we recognize that this lower bound of the coupling strength is conservative, as will be completely removed in the next subsection and will also be demonstrated in the illustrative example.  
\end{rem}

\subsection{Locally Optimal Stabilizing Design for Edge Dynamics}
\label{local-stab}


As seen from (\ref{edge-dyn-2}), the stabilizing design for the transformed edge dynamics is actually decentralized, which means that the stabilizing controller for the transformed edge dynamics can be designed locally, i.e., each edge is equipped with a local performance index as follows, with $i=1,\ldots,N-1$,  
\begin{equation}
	\label{p-index-local}
	J_{i} = \int_{0}^{\infty}{(\tilde{z}_{2,i}^TQ_1\tilde{z}_{2,i}+\tilde{w}_{2,i}^TR_1^{-1}\tilde{w}_{2,i})dt}, 
\end{equation}
where $Q_1 \in \mathbb{R}^{n\times n}$, $Q_1\succeq 0$, and $R_1 \in \mathbb{R}^{m\times m}$, $R_1\succ 0$ are the same as in Theorem \ref{stabilizing-ctlr}. 
Then the controller gain for each edge dynamics is calculated by $K=R_1B^TP_1,$ 
where $P_1 \in \mathbb{R}^{n\times n}$, $P_1\succ 0$ is the solution of the Riccati equation (\ref{Req-local}). 
From LQR theory, it is known that the eigenvalues of $A$ is shifted to the open left half complex plane by the LQR controller with its gain $K$. Therefore, by scaling with a parameter  
$\mu\gamma_i>0 \; \forall \; 1\leq i \leq N-1$, the controller gain $\mu\gamma_iR_1B^TP_1$ still shifts the eigenvalues of $A$ to the left though it could be more or less depending on whether $\mu\gamma_i>1$ or $\mu\gamma_i<1$. Since we have assumed that all eigenvalues of $A$ are on the closed left half complex plane, this means all eigenvalues of $A-\mu\gamma_iBR_1B^TP_1$ belong to the open left half complex plane for all $i=1,\ldots,N-1$. 
Thus, the transformed edge dynamics (\ref{edge-dyn-2})	is stabilized by the locally optimal decentralized controller $\tilde{w}=-\mathcal{F}\tilde{z}$ as in Theorem \ref{ctlr-equi}, and equivalently the initial MAS (\ref{mas}) reaches consensus with the controller (\ref{consensus-ctlr}), for any $\mu>0$.
This gives us the following theorem for the locally optimal distributed consensus design. 
\begin{thm}
\label{consensus-ctlr-local-opt}
For any $\mu>0$, the transformed edge dynamics (\ref{edge-dyn-2})	is stabilized by the locally optimal decentralized controller $\tilde{w}=-\mathcal{F}\tilde{z}$ as in Theorem \ref{ctlr-equi} with the controller gain $K=R_1B^TP_1$ and $P_1$ is the solution of (\ref{Req-local}). Accordingly, the initial MAS (\ref{mas}) reaches consensus with the controller (\ref{consensus-ctlr}) for any 
$\mu>0$. 
\end{thm}

The remarkability of Theorem \ref{consensus-ctlr-local-opt} is that {\it there is no other condition on the coupling strength $\mu$ rather than its positivity}, which to the best of our knowledge has not been proposed hitherto. All existing results, e.g., \cite{Z.Li:2010,Ma:2010,F.Xiao:2007,Movric:2014} required $\mu$ to be greater than some certain bounds. Furthermore, those lower bounds of $\mu$ are dependent on the graph Laplacian $L$, which is a global information and hence adaptive methods, e.g. \cite{Z.Li:2015}, should be designed to overcome this drawback, or other methods should be developed to compute the lower bound of $\mu$ in a distributed way. 
On the other hand, our result shows that no further techniques need to be implemented, which greatly simplifies both the theoretical study and the implementation of the consensus law. 

It is also interesting that the sum of all minimum local performance indexes (\ref{p-index-local}) is $\sum_{i=1}^{N-1}\tilde{z}_{2,i}(0)^TP_1\tilde{z}_{2,i}(0)$, which is the same with that of the global performance index in (\ref{J-global}) since $Q_{1}$ and $R_{1}$ in (\ref{p-index-local}) are equal to those in (\ref{w-matrices}) and hence result in the same $P_{1}$, although the designs are different.

\section{Reduced-Order Distributed Consensus Controller Design }
\label{reduced-design}

\subsection{General Approach}

Let $(\lambda_{1},\ldots,\lambda_{q})$ be eigenvalues of $A$ that are unstable or needed to move further to the left of the imaginary axis and 
$\nu_{1}^{*},\ldots,\nu_{q}^{*}$ are the associated left-eigenvectors; $0<q<n$. We select the weighting matrix $Q_{1}$ under the form
\begin{equation}
	\label{Q1}
	Q_{1}=V \tQ_{1} V^{*},
\end{equation}  
where $V=\begin{bmatrix} \nu_{1} & \cdots & \nu_{q} \end{bmatrix}$, $\tQ_{1} \succeq 0$. 
Denote
\begin{equation}
	\Lambda= \mathrm{diag}\{\lambda_{i}\}_{i=1,\ldots,q}, ~ H=B^T V. 
\end{equation}
We then show in the following theorem that this selection of $Q_1$ in addition to the designed consensus controller gives us a reduced-order consensus controller for the MAS (\ref{mas}).

\begin{thm}
\label{reduced-cctlr}
By choosing the sub-weighting matrix $Q_{1}$ as in (\ref{Q1}) such that $(\tQ^{1/2},I_{N-1}\otimes A)$ is still detectable, the controller (\ref{consensus-ctlr}) designed in Theorem \ref{stabilizing-ctlr} or Theorem \ref{consensus-ctlr-local-opt} has the order of at most $q<n$, resulting in a reduced-order distributed consensus controller for the given MAS (\ref{mas}). Furthermore, the eigenvalue set the closed-loop MAS matrix $\mathbb{A}$ can be characterized as follows,
\begin{equation}
	\label{eigen-set-A}
	\sigma(\mathbb{A})=
	\left(\bigcup_{\gamma \in \sigma(L)}\sigma(\Xi_{\gamma})\right)\bigcup\left(\sigma(A)\backslash \{\lambda_{1},\ldots,\lambda_{q}\}\right),
\end{equation}
where $\Xi_{\gamma}$ is defined by
\begin{equation}
	\label{xi-hetero}
	\Xi_{\gamma}=\Lambda-\mu\gamma H^*R_1H\tP_{1}.
\end{equation}
Simultaneously, the eigenvalue set of the closed loop transformed edge dynamics can be defined by
\begin{equation}
	\label{eigen-set-edge}
	\left(\bigcup_{\gamma \in \sigma(L), \gamma \neq 0}\sigma(\Xi_{\gamma})\right)\bigcup\left(\sigma(A)\backslash \{\lambda_{1},\ldots,\lambda_{q}\}\right).
\end{equation}
\end{thm}

\begin{IEEEproof}
Suppose that $\tP_{1}\in\mathbb{R}^{q\times q}, \tP_{1} \succ 0$ is the unique solution of the following Riccati equation   
\begin{equation}
	\label{R-eq-3}
	\tP_{1}\Lambda+\Lambda\tP_{1}-\tP_{1}\tR_{1}\tP_{1}+\tQ_{1}=0,
\end{equation}
where $\tR_{1}=V^{*}BR_{1}B^TV.$ 
Let $P_{1}=V\tP_{1} V^{*}, P_1\succ 0$, then substituting $Q_{1}$ and $P_{1}$ back to the Riccati equation (\ref{Req-local}), we can easily verify that $P_1$ is a solution of (\ref{Req-local}). 
Since (\ref{Req-local}) has a unique positive semidefinite solution, $P_1=V\tP_{1} V^{*}$ is indeed that unique one. 
Since $\mathrm{rank}(R_{1}B^TP_{1})=\mathrm{rank}(R_{1}B^TV\tP_{1} V^{*}) \leq q$, the controller designed in Theorem \ref{stabilizing-ctlr} becomes a reduced-order distributed consensus controller whose order is at most $q$.  

Next, we will show that the eigenvalue spectrum of the closed-loop matrix $\mathbb{A}$ can be characterized by (\ref{eigen-set-A}). 
Consider any right eigenvector $\eta_j\in\mathbb{C}^{n}$ of $A$ corresponding to an eigenvalue 
$\lambda_j\in\sigma(A)\backslash\{\lambda_{1},\ldots,\lambda_{q}\}$, $j=q+1,\ldots,n$. 
For $i=1,\ldots,N$, denote $e_i\in\mathbb{R}^{N}$ the vector whose $i$th element is $1$ while all other elements are zero. Then
\begin{align}
	\label{eig-eq1}
	\mathbb{A}(e_i \otimes \eta_j) =& e_i\otimes(A\eta_j)-\mu(Le_i)\otimes(BR_{1}B^TP_{1}\eta_j), \nonumber \\
	=& \lambda_j(e_i\otimes \eta_j)-\mu(Le_i)\otimes(BR_{1}B^TV\tP_{1} V^{*}\eta_j), \nonumber \\
	=& \lambda_j(e_i\otimes \eta_j),
\end{align}
due to a fact that $V^{*}\eta_j=0$ since $\nu_{l}^{*}\eta_j=0 \;\forall\; l=1,\ldots,q.$ 
Therefore, $\lambda_j$ is an eigenvalue of $\mathbb{A}$ with the associated right eigenvector $e_i\otimes \eta_j$. This leads to
\begin{equation}
	\label{eig-eq2}
	\left(\sigma(A)\backslash \{\lambda_{1},\ldots,\lambda_{q}\}\right) \subset \sigma(\mathbb{A}),
\end{equation} 
with the note that each eigenvalue $\lambda_j,j=q+1,\ldots,n$ has the multiplicity $N$. 
On the other hand, 
\begin{align*}
	(I_N\otimes V^{*})\mathbb{A} &= (I_N\otimes V^{*})[I_{N}\otimes A-\mu L\otimes(BR_{1}B^TP_1)] \\
	&= I_N\otimes (\Lambda V^{*})-\mu L\otimes(V^{*}BR_{1}B^TV\tP_{1} V^{*}),  \\
	&= [I_N\otimes \Lambda - \mu L\otimes(H^{*}R_{1}H\tP_{1})](I_N\otimes V^{*}). 
\end{align*}
Let $\rho^{*}$ be any left eigenvector of $L$ and $\gamma$ is the associated eigenvalue. Then multiplying both sides of above equation with 
$\rho^{*}\otimes \xi^{*}$, $\xi \in \mathbb{C}^{q}$, we obtain
\begin{align}
	\label{eig-eq4}
	& (\rho^{*}\otimes \xi^{*})(I_N\otimes V^{*})\mathbb{A} \nonumber \\
	=& (\rho^{*}\otimes \xi^{*})[I_N\otimes \Lambda - \mu L\otimes(H^{*}R_{1}H\tP_{1})](I_N\otimes V^{*}), \nonumber \\
	=& [\rho^{*}\otimes(\xi^{*}\Lambda)-\mu\gamma\rho^{*}\otimes(\xi^{*}H^{*}R_{1}H\tP_{1})](I_N\otimes V^{*}),  \nonumber \\
	=& \rho^{*}\otimes(\xi^{*}\Lambda-\mu\gamma \xi^{*}H^{*}R_{1}H\tP_{1})(I_N\otimes V^{*}).
\end{align}
Now, suppose that $\xi^{*}$ is a left eigenvector of $\Xi_{\gamma}$ defined in (\ref{xi-hetero}) and the corresponding eigenvalue is $\alpha$, it is deduced from (\ref{eig-eq4}) that
\begin{align}
	\label{eig-eq5}
	(\rho^{*}\otimes \xi^{*})(I_N\otimes V^{*})\mathbb{A} &= \rho^{*}\otimes(\xi^{*}\Xi_{\gamma})(I_N\otimes V^{*}),  \nonumber \\
	&= \alpha(\rho^{*}\otimes\xi^{*})(I_N\otimes V^{*}),  \nonumber \\
	&= \alpha\rho^{*}\otimes(\xi^{*}V^{*}).
\end{align}
This means $\alpha$ is also an eigenvalue of $\mathbb{A}$ with the associated left eigenvector $\rho^{*}\otimes(\xi^{*}V^{*})$. Furthermore, $\alpha$ has the multiplicity $N$. Accordingly,
\begin{equation}
	\label{eig-eq6}
	\sigma(\Xi_{\gamma}) \subset \sigma(\mathbb{A}).
\end{equation}
Thus, combining (\ref{eig-eq2}) and (\ref{eig-eq6}) gives us (\ref{eigen-set-A}). 

The eigenvalues set of the closed-loop transformed edge dynamics can be proved in a similar manner. 
\end{IEEEproof}

\subsection{First-order Distributed Consensus Controller}

Assuming that $A$ has a single $0$ eigenvalue. From the result of Theorem \ref{reduced-cctlr}, we see that to ensure the consensus of the given MAS or the stability of the transformed edge dynamics, 
the reduced-order consensus controller should shift this $0$ eigenvalue of $A$ to the left haft complex plane. 
Let $\nu^T\in \mathbb{R}^{1\times n}$ be the left eigenvector of $A$ associated with the eigenvalue $0$. Consequently, choose $Q_1=\nu q_1\nu^T$ where $q_1 > 0$. 
Suppose that this selection of $Q_1$ still guarantees that $(\tQ^{1/2},I_{N-1}\otimes A)$ is detectable. 
The following result is immediately obtained as a corollary of Theorem \ref{reduced-cctlr}.

\begin{coro}
\label{coro-1}
The MAS (\ref{mas}) achieves consensus by a $1$st-order distributed controller $u=-\mu (L \otimes K)x$ with any $\mu>0$ and 
\begin{equation}
	\label{K-1st-order}
	K=\frac{\sqrt{q_{1}}}{\sqrt{r_{1}}}R_1B^Tvv^T,
\end{equation}
where $r_1=v^TBR_1B^Tv$. The closed-loop eigenvalues of the MAS are 
\begin{equation}
	\left(\bigcup_{\gamma \in \sigma(L)}-\mu\gamma\sqrt{q_{1}r_{1}}\right)\bigcup\left(\sigma(A)\backslash \{0\}\right),
\end{equation}
and of the transformed edge dynamics are
\begin{equation}
	\label{edge-eigen}
	\left(\bigcup_{\gamma \in \sigma(L),\gamma \neq 0}-\mu\gamma\sqrt{q_{1}r_{1}}\right)\bigcup\left(\sigma(A)\backslash \{0\}\right).
\end{equation}
Furthermore, the consensus speed is determined by
\begin{equation}
	\min\{\mu\sqrt{q_{1}r_{1}}\lambda_{\min}(L),\lambda_{\min}(-A)\}.
\end{equation}
\end{coro}

The strong advantages of Corollary \ref{coro-1} are as follows. First, it gives us a superior simplicity for distributed consensus controller design since the controller's order is 
just $1$. Second, it clearly supports our relaxation on the coupling strength $\mu>0$ in Theorem \ref{consensus-ctlr-local-opt} by looking at the eigenvalue set (\ref{edge-eigen}) of the transformed edge dynamics. And last, the consensus speed and the eigenvalue distribution of the closed-loop MAS are explicitly established, whereas they are implicit in the full-order distributed consensus controller design. In addition, the consensus speed is adjustable by varying $\mu$, $q_{1}$, and $r_{1}$. 

\begin{rem}
Although only the results of $1$st-order distributed consensus controllers are presented in this section, a similar design is readily obtained for $2$nd-order distributed consensus controllers in the scenario that $A$ has a pair of complex conjugate eigenvalues on the imaginary axis. Particularly, $Q_1$ is chosen as in (\ref{Q1}) with $V$ composes of two left eigenvectors associated with the eigenvalues on the imaginary axis of $A$. 
\end{rem}

\subsection{Illustrative example }

\begin{figure}[ht!]
		\centering
		\includegraphics[scale=0.4]{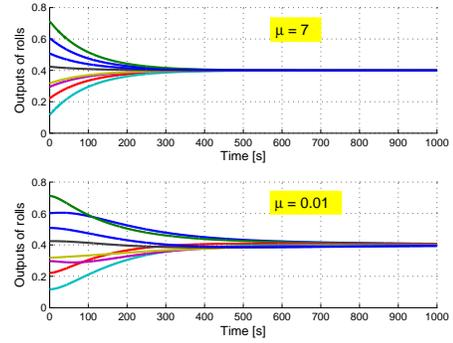}
		\caption{Consensus in the drying section of a paper converting machine by a $1$st-order consensus controller with different values of the coupling strength.}
		\label{pmachine_compare}
	\end{figure}

Consider a drying section of a paper converting machine including of $9$ rolls \cite{Nguyen:2015j} where the control target is to synchronize the angles of the rolls in order to make the paper run smoothly without tearing. Assuming here that each roll just communicates with the rolls in front of and behind it. 
The state matrices of the $k$th roll, $k=1,\ldots,9,$ is 
\begin{equation*}
	A=\begin{bmatrix} 0 & 1 & 0 \\ 0 & -0.01 & 0.2 \\ 0 & 0 & -125 \end{bmatrix}, 
	B=\begin{bmatrix} 0 \\ 0 \\ 20 \end{bmatrix},
	C=\begin{bmatrix} 1 \\ 0 \\ 0 \end{bmatrix}^T.  
\end{equation*}
It can be seen that $A$ has a single $0$ eigenvalue while the other two eigenvalues belong to $\mathbb{C}_{-}$. Therefore, we apply the $1$st-order consensus controller design in Corollary 
\ref{coro-1} with $R_1=100$ and $q_1=1$. Figure \ref{pmachine_compare} displays the consensus of the rolls' outputs as $\mu=7$ for globally optimal design, and $\mu=0.01$ for locally optimal design. 
We can observe that even the coupling strength $\mu$ is very small but positive, the consensus is still achieved. This confirms our result on the relaxation of the conservative condition for the coupling strength $\mu$. 

\section{Conclusion}
\label{concl}

This paper has proposed a systematic approach to design distributed consensus controllers for general linear MASs with the following remarkable features. 
First, a transformation has been introduced, based on which the distributed consensus design is shown to be equivalent to a distributed stabilizing design for the transformed system called edge dynamics. 
Then the conservative bound for the coupling strength is completely removed by the locally optimal synthesis, resulting in a fully distributed consensus controller. Second, a method has been presented to reduce the consensus controller's order which can be $1$ if the agents have a single pole at the origin. This gives us not only a superior simplicity for the distributed consensus synthesis, but also an explicit description of consensus speed and eigenvalue distribution of the MAS.


%


\section*{Acknowledgment}

The author would like to thank the anonymous reviewers for their valuable comments and suggestions, 
and to send special thanks to Toyota Technological Institute for its supports.

\ifCLASSOPTIONcaptionsoff
  \newpage
\fi



%

\bibliographystyle{IEEEtran}
\bibliography{IEEEabrv,References}

%
%

%

%
%
%
%
%
%
%
%
%
\end{document}